# Ethics Understanding Of Software Professional In Risk Reducing Reusability Coding Using Inclusion Set Theory

G.Singaravel
Department of Computer Science and Engineering
K.S.R.College of Engineering
Tiruchengode, Namakkal,
Tamilnadu, India

Dr.V.Palanisamy
Principal
Info Institute of Engineering
Coimbatore
Tamil Nadu
India

Dr.A.Krishnan
Dean
K.S.R.College of Engineering
Tiruchengode, Namakkal,
Tamilnadu,
India

*Abstract*— The technical skill or ability of an individual is different to person in software developments of projects . So, it is necessary to identify the talent and attitude of an individual contribution can be uniformly distributed to the different phases of software development cycle. The line of code analysis metrics to understanding the various skills of the programmers in code development. By using the inclusion set theory of n (AUB) refer to strength and risk free code developed from union of software professionals and system must comprise of achievement of the system goal, effective memory utilization and intime delivery of the product.

## I. INTRODUCTION

A Software professional has always been a liable person to the successful development of the software. Identifying human resources is more difficult than identifying the software or hardware assets. The concern people with enough knowledge and experience should be assigned to the task. The Engineers and the scientists share a basic drive to accomplish to something which they can point to as their own unique achievement because of the professional need for unique achievement[6],[8]. It is very necessary to match professional with their work assignments. More technical issue and interpersonal skill are very much involved in it. The development team face many challenges and hardships as they are pressurized by the obstacles during development to get the software product out of the door[10].

The best result can be obtained only if the allotment of the different tasks among different software professionals according to their nature and ability[6]. So it is important to identify the unique skill of individual professional and make use of the same for software development process in which result are very less risk level.

**Corresponding Author**
G.Singaravel,Email:singaravelg@gmail.com
Mobile:09943455245,Fax:04288274757

## II. IDENTIFICATION OF TECHNICAL PROFESSIONAL FOR SOFTWARE DEVELOPMENT

Among highly motivated professionals, the most talented generally do the best work. Every field of specialization has a unique set of talents who are responsible for success. The Professional can generally succeed as long as their talents are reasonably consistent with the needs of their work. The talented people are the most important asset of an organization. They originate the creative ideas and solve the key problems to produce the most successful products.

## III. LINE CODE ANALYSIS

This analysis is to identify the risk in perspective of the programmer's knowledge. This is the comparative study to compare the programming skills of different programmers under a particular scenario. Every developer has his own idea to develop a program. The style of program writing of a programmer may be entirely different from other. Hence this study promotes an idea to choose the best programmer among a crew of programmers based on three levels of the program writing.There are Low level program, Medium level program and High level program

**LOW LEVEL PROGRAM:** Low level program is a simple program with limited number of variables. For example, Calculating the area of Rectangle.

**MEDIUM LEVEL PROGRAM:** Medium Level Program is of variables at multilevel systems. For example, Use of Inheritance to access the variable from base class.

**HIGH LEVEL PROGRAM:** High Level Program is involved with complex data variables in the system which depicts the importance of the variable and its scope within the system. For example, Functions used for online quiz system.

The following calculation metrics of LOC,PROGRAM VOLUME ,COMPILATION DETYAIL AND ERROR are support for analysing of three levels program

**LOC (N):** LOC represent Line of Code. LOC is the total number of lines present in a program. If the LOC is a larger value, then it is observed to have greater complexity[1].

**PROGRAM VOLUME(V):** Program volume can be represented by the Eq.(1)

$$V = N \log n \qquad (1)$$

Where,**N**- Total number of lines present in a program,**n**- Sum of the total number of operators and operands present in a program.If program volume is larger, it seems to have much complexity because there must exist many operators and operands[1].

**COMPILATION DETAILS:** Compilation details is based on the program error. This error is divided and analyzed into various categories.**ERROR:** Types of error[1],[7] is shown in figure 2.1

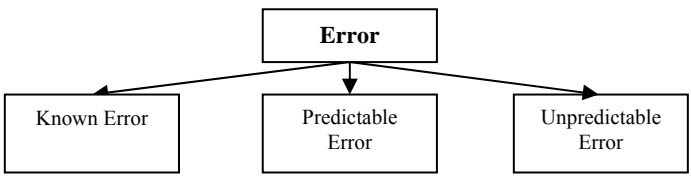

Figure 2.1: Types of Error

1. Known Errors:Errors where the programmers know the exact key to overcome it. Such error may be syntax error or such relative errors.

2. Predictable Errors:These errors are one where the programmers do not know the exact key to the error but they'll predict some solutions. For example, logical errors.

3. Unpredictable Errors:The programmers cannot find a solution to particular error problems. Such errors are known to be unpredictable error. For example, some linker errors which may be implicit or beyond the knowledge of the programmer.

The analysis of the program based on the levels of programming .It is taken as a sample from different levels of coding by different programmers. The data are collected from the experiments conducted at laboratory using various programs.

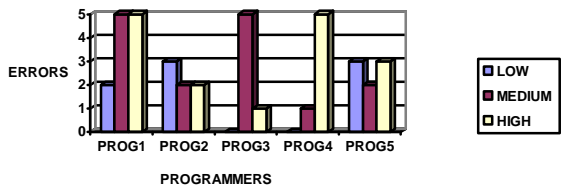

Figure 2.2: Graphical Representation of Programmers Vs Errors

This bar chart analyzes the comparative study for the programming skills of different programmers in different aspect of human ethics of coding development. It also depicts the comparison of various capabilities of the programmers who are all working in different platforms.It shows that the capability level of the programmer differs from each other in figure 2.2.

IV. DICTORIAL RISK FACTOR FOR SOFTWARE PROJECT CODING

**Risk attack in software projects**

The area of highest risk attack slowly starts from analysis phase and goes through design, implementation and ends in testing phase.

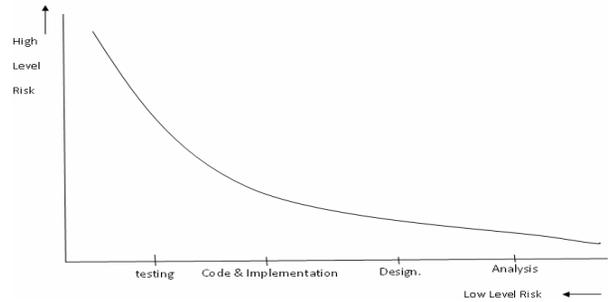

Figure 2.3: Risk Level during the development phases

This figure2.1 show the area of the project where the project manager has high level of risk at implementation and coding phase on wards[3],[7].As long as risks are determined and securely mitigated, the quality of risk will slowly move on its predicted path. If risks are not identified and are not securely mitigated, the bugs will rapidly increase in implementation phase and finally the software project fails to satisfy the requirements of the customer[4].

Risk due to inadequate knowledge

Even though software developing team identifies the risk from design phase, they have to overcome many risks occurred at coding phase also. Professional who is of lack of knowledge on particular area may tend to increase the risk of the product. The primary goal of a software development team is to develop code and documentation that will meet the project's requirement. The primary issue is the software must be maintainable and reusable. According to the impact level of risk occurred in the system, that can be classify into three basic Risk level Identification are Low Level,Medium Level and High Level.

Consider the Table 2.5 of risk controlling factors which depicts the various factors which influence risk in the system through percentage of risk level identified in Eq.(2).

$$\text{Percentage of Risks Level Identified in the system} = \frac{\text{No.of Risk Factors which produce major effect in the system}}{\text{Total No.of Risk Factors in the system}} \times 100\,\%$$

The Risk Level controlling factors into three .There are Low Percentage of risk lies between 0 % to 30 % , Medium Percentage of risk lies between 31 % to 60 % ,High Percentage of risk lies between 61 % to 100%.

TABLE 2.5 RISK CONTROLING FACTORS

| Factors | Example | Risk Level |
|---------|---------|------------|
| Error | Locate error | Low |
|  | Analyze error | Low |
|  | Estimate error | medium |
| Bugs | Control bugs | Medium |
|  | Runtime bugs | Low |
|  | Software bugs | Medium |
|  | Unauthorized access | high |
| Faults | Wrong boundary value | Medium |
|  | Initialization problem | Low |
|  | Reference | Low |
|  | Format inconsistency | Low |
| Failures | Transient failure | High |
|  | Unrecoverable | High |
|  | computing | Low |

The component invocation bugs, problem with parameters, divided by zero are sample examples which depict the various factors which are affecting the implemetation coding in software projects risk due to inadequate knowledge of programmers[2].

**Component invocation bugs**: The wrong component ie a faulty one or a non existing component is invoked. For example, program written by C++

```
class count {
public static void main (string args [])
{
…
…
for (i=0;i<10;i++)
{
int count ;
count ++;
}
system.out.println (count);
}
```

This program shows an error: **unable to find symbol count**. The error is because the scope of count is only inside the inner loop, so the component is invoked as wrong place.

**Problem with parameters**: Parameters may be incorrect in their number, order data type, value, connections etc. For example, program written by C++

```
class rectangle
{
…
…
rectangle (int x, int y)
{
i =x;
w = y;
}
}
class area
{
public static void main (arg[])
{
rectangle rect1 = new rectangle (15);
…
…
}
```

// **error type mismatch in number of argument.**

Here a two argument constructor is defined by a single argument constructor is called. Hence type mismatch in number of argument passed occurs.

**Divided by zero** : If any arithmetic expression results in a denominator value of 0 then an error occurs. For example, class Test

```
{
public static void main(String args[])
{
int a = 5;
int b = 2;
int c = 3;
a = a/( c-b-1);
}
}
```

The expression result in a denominators of zero and hence error result.

### V. TWO LEVELS OF SOFTWARE PROFESSIONALS

Software comprising of several modules are developed by a team of people in a software company. Each module is developed by an individual or a group of people it is well known that risks to the software depends on various factor such as skill, requirements, scheduling and cost[11]. Among those factors skill of an individual on a particular language with which the product being developed gives a greater impact on risk of the software. In general, Software professional have been classified as Vertical Software Professional (VSP) and Horizontal Software Professional (HSP) in figure 2.4. Vertical Software Professionals(VSP) are those who are expert in a particular language and they can't do well in the other languages except the one at which they are very potential. Horizontal Software Professional (HSP)are those who are not expert in a particular language and they can be also moderately in the other languages. For example, Professionals from group 'A' are expert in "C" language, they are capable of solving all problems in "C" language. Professionals from group 'B' are not expert at all software developing language but they know something of all those developing language. They tend to commit mistake in developing cycle of the software.

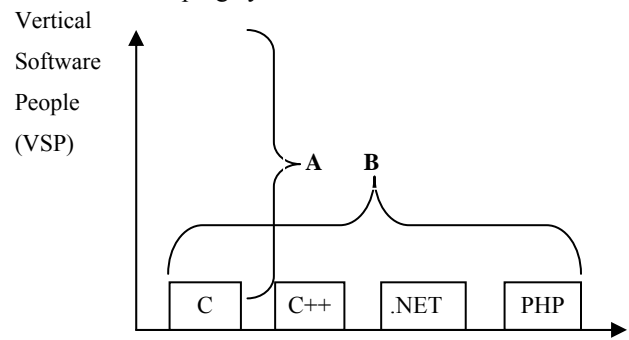

Figure 2.4 VSP vs HSP

## VI. ALLOCATION OF SOFTWARE PROFESSIONALS USING SET THEORY

A fundamental concept of set theory is that of membership or belonging to set, any object belongs to a set a member or an element of that set. The object in sets may be anything, example numbers, people, rivers, car, and mountains. If an object x is a member of a set A, then write as x E A which may be read as "x belongs to A" in other word "x is an element of the set A", in vice versa x not belong to A when object x is not a member of the set A[9]. So software people belonging each other for make a projects in way of vertical software people belongs to horizontal software people, horizontal software people belongs to vertical software people.

Discussed the notation of membership of an element in a set. Another basic concept in set theory is that of inclusion. Let A and B be any two sets. If every element of A is an element of B, then A is called a subset of B, or A is said to be included in B, or B includes A. Symbolically, this relation is denoted Eq.(3).

$$A \subseteq B \Leftrightarrow (x)(x \in A \rightarrow x \in B) \Leftrightarrow B \supseteq A \quad (3)$$

The Mathematical abstraction below will show how the risk reduction possible in developing a software.
Let as, set 'A' be the VSP and set 'B' be the HSP. Each set A and B has N number of people is term denoted in Eq.(4).

$$A = \{A1, A2, A3, \ldots \ldots N\}, B = \{B1, B2, B3, \ldots \ldots N\} \quad (4)$$

Professionals who belong to set A are expert at any one of the developing resources of the software like C, C++, JAVA etc,. These professionals have been only under particular platform at which they are said to be expert. Professionals who belong to set B are not expert in any specific developing tools of resource but they have gained considerable knowledge in almost all language as they move from one platform to another dynamically.

By using venn diagram of inclusion set theory in given Eq.(5).
$$n(A \cup B) = n(A) + n(B) - n(A \cap B) \quad (5)$$
where n (AUB) refer to strength or risk free code developed from union of VSP and HSP, n (A) refers to total skill of VSP, n (B) refers to total skill of HSP, n (A∩B) refers to risk reduced by the combination of the groups when they are employed together on a project in figure 2.5.

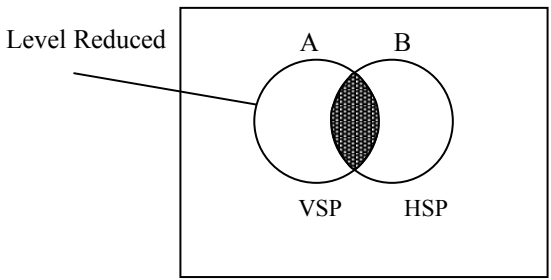

Figure 2.5 Risk Level Reductions by combining VSP and HSP

When both groups of VSP and HSP are combined together for a project, the chain of risk can be considerably reduced. If A1 and B1 are disjoint sets, then the risk level is very high. If $A_1$ and $B_1$ are intersected, then the risk level can be considerably reduced by using inclusion set theory of Risk reduction chain in Eq.(6).of VSP and HSP and also is given diagrammatically representation in figure 2.6.
$(A_1 \cap B_1)+(B_1 \cap A_2)+(A_2 \cap B_2)+(B_2 \cap A_3)+(A_3 \cap B_3)+\ldots+(A_N \cap B_N)$

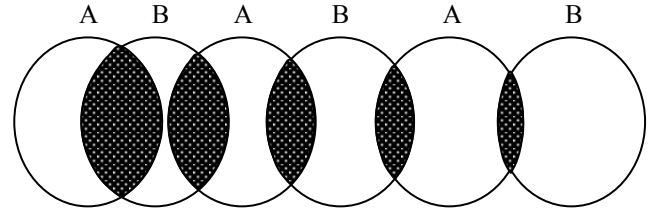

Figure 2.6 Risk reduction chain of VSP and HSP

## VII. APPLICATION OF THE PROPERTIES OF SET THEORY FOR VSP AND HSP.

Let us take as A= VSP, B= HSP.
If the inclusion set theory property is satisfy with mathematical notation using VSP and HSP, the satisfy level is LOW RISK. If the set theory property is not satisfy, then level is HIGH RISK.
Table 2.5: Usage of Set Theory in assigning the software professionals for software development

## VIII. RESULT AND DISCUSSION.

The line of code analysis results in understanding the various skills of the programmers in code development. It also represents the system growth with various levels of programmers involved in the development of the system. The following graph figure 2.8 depicts the efficiency of risk level when we use different people (HSP and VSP) for development. It shows the risk level gets reduced when we combine both the HSP and VSP people. It satisfies the set theory relationship based on the HSP and VSP people programming behaviours.

| S.No | Property | Syntax | Explanation | Level of risk |
|---|---|---|---|---|
| 1. | Inclusion of Sets | $A \subseteq B \Leftrightarrow$ (x) (x$\in$ A$\rightarrow$ x $\in$B) $\Leftrightarrow B \subseteq A$ | VSP is included in HSP, or HSP is included in VSP. | Very low risk |
| 2. | Equality of Sets | $A = B \Leftrightarrow$ ($A \subseteq B \wedge B \subseteq A$) | VSP and HSP are equal in number. | Low risk |
| 3. | Intersection of Sets | $A \cap B =$ {x \| (x E A ) $\wedge$ (x E B)} | The intersection of any two sets VSP and HSP, written as VSP $\cap$ HSP is the set consisting of professionals who are of same category of skills both in VSP and HSP. | Low risk |
| 4. | Disjoint sets | $A \cap B = \phi$ | VSP and HSP have no talent in common | Very high risk |

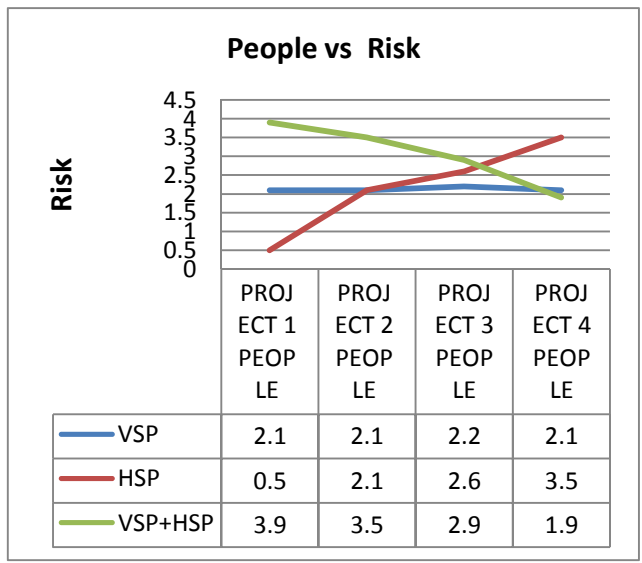

Figure 2.7 People Vs Risk

The risk level of the system can be reduced considerably when we use the HSP and VSP professional together. It also depicts that the system must comprise of both these professionals for effective achievement of the system goal. In order to make a system efficient, make use of the reusable components which results in the reduction of development time, effective memory utilization and intime delivery of the product.

## VIII. CONCLUSION

The Talented software professionals are great asset to an organization to avoid the risk factors. Analysis for Line of codes and performances of different software professionals are different for the same program undertaken. An Organization can allot the works to the professionals of mixed experiences to reduce the risk factor and time schedule.